\documentclass{PoS}

\usepackage{amsmath}
\usepackage[switch]{lineno}
\usepackage{lipsum}
\newcommand{\aj}{AJ} 
\newcommand{\apj}{ApJ} 
\newcommand{\prd}{Phys. Rev. D } 
\newcommand{\aap}{A\&A} 
\newcommand{\mnras}{MNRAS} 
\newcommand{\araa}{ARA\&A } 

\title{Energy-dependent morphology of the pulsar wind nebula HESS\,J1825-137 seen by \textit{Fermi}-LAT}

\ShortTitle{Energy-dependent morphology of the PWN HESS\,J1825-137 seen by \textit{Fermi}-LAT}

\author{\speaker{G. Principe}\thanks{On behalf of the \textit{Fermi}-LAT collaboration.}\\
        INAF - Istituto di Radioastronomia, via Gobetti 101, I-40129
 Bologna, Italy\\
        E-mail: \email{giacomo.principe@inaf.it}}

\author{A.M.W. Mitchell\\
        Physik-Institut, Universit\"at Z\"urich, Winterthurerstrasse 190, CH-8057 Z\"urich Switzerland.}
\author{J. Hinton\\
        Max-Plank-Institut f\"ur Kernphysik, Saupfercheckweg 1, D-69117 Heidelberg, Germany}
\author{R.D. Parsons\\
        Max-Plank-Institut f\"ur Kernphysik, Saupfercheckweg 1, D-69117 Heidelberg, Germany}
\author{S. Caroff\\
         Sorbonne Universit\'e, Universit\'e Paris Diderot, Sorbonne Paris Cit\'e, CNRS/IN2P3, Laboratoire de Physique Nucl\'eaire et de Hautes Energies, LPNHE, 4 Place Jussieu, F-75252 Paris, France}
\author{J. Hahn\\
        Max-Plank-Institut f\"ur Kernphysik, Saupfercheckweg 1, D-69117 Heidelberg, Germany}
\author{S. Funk\\
        Erlangen Centre for Astroparticle Physics, Erwin Rommel Strasse 1, D-91058 Erlangen, Germany}

\abstract{Taking advantage of 10 years of \textit{Fermi}-LAT data, we perform a new and deep analysis of the pulsar wind nebula (PWN) HESS\,J1825-137. We present the results of the spectral analysis and of the first energy-resolved morphological study of the PWN HESS\,J1825-137 from 1 GeV to 1 TeV. This PWN is an archetypal system making it a perfect laboratory for studying particle transport mechanisms. Combining this analysis with recent HESS results enables a more complete picture of the nebula to emerge.}

\FullConference{36th International Cosmic Ray Conference -ICRC2019-\\
		July 24th - August 1st, 2019\\
		Madison, WI, U.S.A.}

\begin{document}

\section{Introduction}
\label{sec:intro}

The core collapse of a massive star may create a rapidly rotating and magnetized neutron star (pulsar), which can emit a wind of relativistic, 
charged particles with luminosity around $10^{38}$ erg\,s$^{-1}$. Confinement of these outflows generates luminous pulsar wind nebula (PWN). The composite supernova remnant (SNR) is charcterized by the evolution of this PWN inside the interior region of the collapsed star\cite{2017ApJ...844....1K}. While the majority of these objects maintain a spherical symmetry during the early-stage evolution \cite{2006ARA&A..44...17G}, the composite system is particularly interesting when its relic PWN tends to form complex structure during the late-stage devolopment.
Evolved PWNe, due to their large $\gamma$-ray extension and possible energy-dependent morphology are ideal candidates for investigating particle transport mechanisms inside celestial objects. 

Among such objects, HESS\,J1825-137 is the largest and one of the most TeV efficient $\gamma$-ray emitting PWN currently known \cite{2019A&A...621A.116H,2018A&A...612A...2H}. It is powered by a young and very energetic pulsar PSR\,J1826-1334 (also known as PSR\,B1823-13), which was discovered in 1992  \cite{1992MNRAS.254..177C}.
The pulsar has characteristics very similar to the Vela pulsar: it has a spin period of 101.48\,ms, a characteristic age of 21\,kyr, a spin-down energy of 2.8 $\times 10^{38}$ erg s$^{-1}$ and it is situated at a distance of 3.9$\pm$0.4 kpc \cite{2005AJ....129.1993M}. 
The PWN HESS\,J1825-137 presents an asymmetric emission, towards the south of the pulsar location, which has been suggested to be caused by the interaction of the progenitor SNR with molecular clouds north of the pulsar leading to a reverse shock interaction early in the system evolution. 
The discovery of the energy-dependent morphology of the PWN HESS\,J1825-137 at TeV energies \cite{2006A&A...460..365A} indicates that the emission is dominated by `relic' electrons from the earlier epochs of the nebula in which the pulsar was spinning down more rapidly, therefore releasing more energy into the system. 

In the GeV regime, the source was first detected in 2011 by the Large Area Telescope (LAT) \cite{2009ApJ...697.1071A}, on board the \textit{Fermi Gamma-ray Space Telescope}.
Previous LAT analyses of the PWN HESS\,J1825-137 have been performed using 20 months of data in the 1 -- 100\,GeV energy band \cite{2011ApJ...738...42G} and subsequently 6 years of data in the 10\,GeV -- 1\,TeV energy band \cite{2017ApJ...843..139A}. Despite the LAT energy threshold of 20 MeV, the source is not resolved in the MeV range  \cite{2018A&A...618A..22P}. 

Taking advantage of the large amount of available \textit{Fermi}-LAT data, an analysis of the energy-resolved morphology and spectral parameters of this source is performed using 10 years of LAT data in the energy range between 1\,GeV and 1\,TeV.

\section{\textit{Fermi} data and analysis}
\label{sec:data}

For the analysis, we use 10 years (between August 4, 2008 and August 3, 2018) of Pass 8 Source class events \cite{2013arXiv1303.3514A} in the energy range between 1\,GeV and 1\,TeV.  The data are taken in a region of interest (ROI) of radius 15$^{\circ}$ and centered on the PWN position given in \cite{2017ApJ...843..139A}. The P8R2\_Source\_V6 instrument response functions (IRFs) are used. In order to eliminate most of the contamination from secondary $\gamma$-rays from the Earth's limb \cite{2009PhRvD..80l2004A}, we exclude $\gamma$-rays with zenith angle larger than 105$^{\circ}$.

The analysis is performed with Fermipy\footnote{http://fermipy.readthedocs.io/en/latest/} \cite{2017ICRC...35..824W}, a python package that facilitates analysis of data from the LAT with the \textit{Fermi} Science Tools, of which the version 11-07-00 is used.
We estimate the general parameters of the source (localization, averaged extension and spectra) using the complete energy range between 1\,GeV and 1\,TeV. 
The model used to describe the sky in this ROI includes all point-like and extended LAT sources within 20$^{\circ}$ from the PWN position listed in the FL8Y\footnote{https://fermi.gsfc.nasa.gov/ssc/data/access/lat/fl8y/gll\_psc\_8year\_v5.fit} list, as well as the Galactic diffuse \cite{2017ApJ...840...43A} and isotropic emission.
We model the source using a 2D-Gaussian model for the spatial template and a LogParabola spectral model $\left(\frac{dN}{dE} = N_{0}(\frac{E}{E_{0}})^{- [\alpha + \beta log(E/E_{0})]}\right)$ \cite{2017ApJ...843..139A}. 


\section{The PWN HESS J1825-137}

\subsection{Localization and averaged extension analysis}

\begin{figure}[h]
\centering
\includegraphics[scale=0.69]{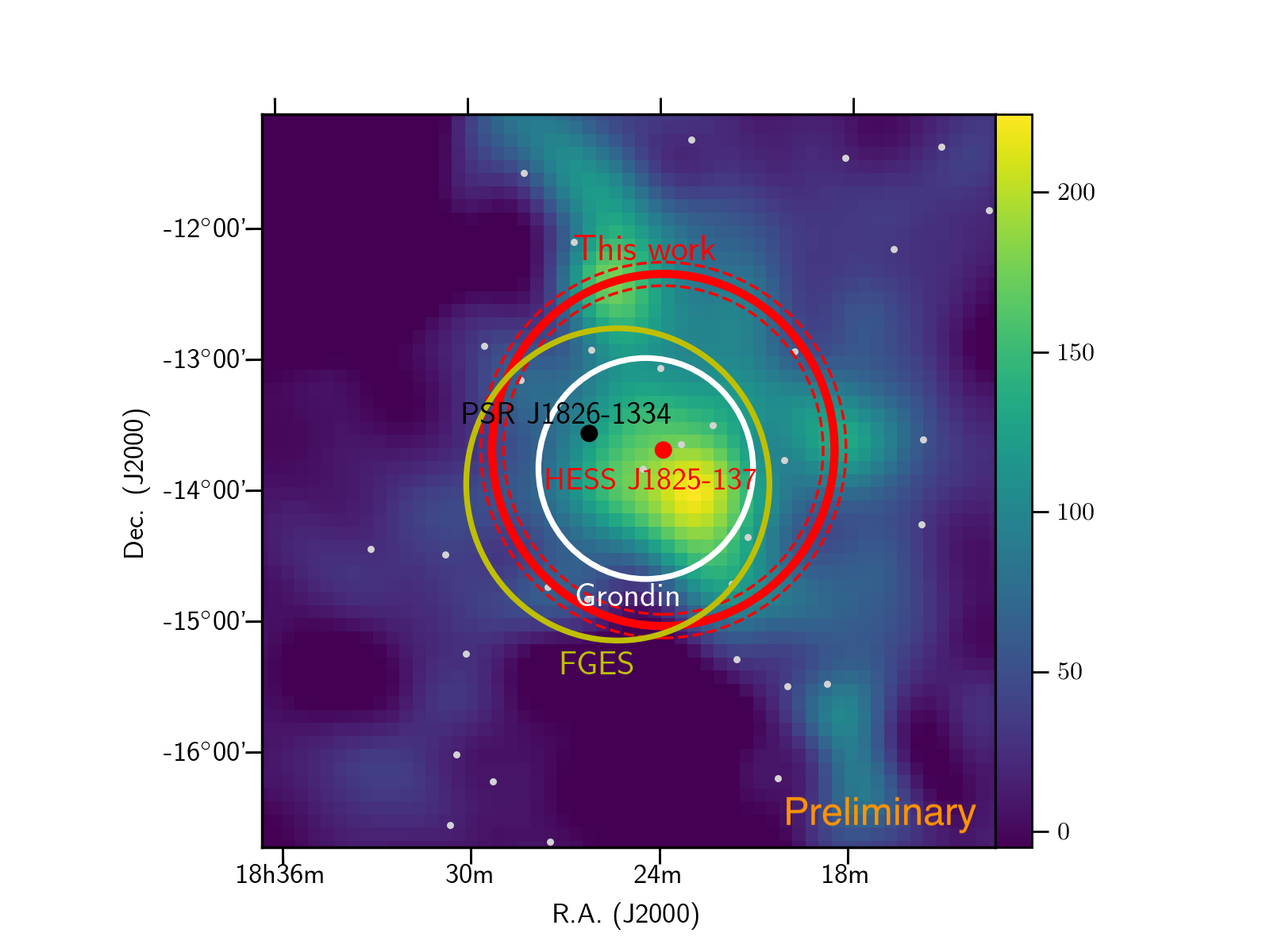}
\caption{\small \label{counts_map} Excess map, in celestial coordinates, of the region around the PWN HESS\,J1825-137 for the energy range between 3\,GeV and 1\,TeV. The red circle and point indicate the 2D-Gaussian extension and centroid fit obtained in this work. The red dashed circles mark the uncertainty on the 2D-Gaussian extension. The green and white circles corresponds to the extension obtained respectively in the FGES catalog \cite{2017ApJ...843..139A} and in \cite{2011ApJ...738...42G}. The black point indicates the position of PSR\,J1826-1334, the pulsar which is believed to power the nebula. The FL8Y sources are represented with light grey points. 
}
\end{figure}

After a preliminary optimization of the parameters of the sources included in the model, we perform the localization and extension analysis first in the entire energy range: 1\,GeV--1\,TeV. 
The source is not significantly detected below 3 GeV, therefore figure \ref{counts_map} shows the excess map of the region around the PWN for the energy range between 3\,GeV and 1\,TeV. 
In the map, the size obtained for the PWN in this work is compared with the radius obtained in the previous analyses with \textit{Fermi}-LAT data.

Table \ref{table_loc} reports the results of the localization and extension analysis performed in the energy range between 1\,GeV and 1\,TeV using a 2D-Gaussian model as spatial template for the source. The extension result reported here corresponds to the 68\% containment radius.

\begin{table}[h]
\small
\centering
\begin{tabular}{cc}
Position & Value \\
\hline
\hline
RA & $275.97^\circ \pm 0.03^\circ$ \\ 
DEC & $-13.71^\circ \pm 0.04^\circ$ \\
Extension $R_{68\%}$ (1\,GeV-1\,TeV) & $1.35^\circ \pm 0.09^\circ $\\
Test Statistic (TS) & 992\\
\hline
\end{tabular}
\caption{\small \label{table_loc}
Localization and extension results. The PWN position corresponds to RA 18h23m52s$\pm$8s and DEC $-13^{\circ}42'36"\pm2'14"$. The results are obtained using 10 years of \textit{Fermi}-LAT data in the energy range between 1\,GeV and 1\,TeV.}
\end{table}

The resulting center position of the PWN is shifted by about 0.58$^\circ$ from PSR\,J1826-1334, the pulsar associated with the nebula. 
The interaction of the progenitor SNR with dense molecular clouds north east of the pulsar, leading to a relatively fast formation of a reverse shock on the northern side of the nebula, has been supposed to be the probable cause of the asymmetry. EVLA observation at 1.4 GHz made by \cite{2012BAAA...55..179C}  reveal, in fact, a nearby molecular cloud with a density of $\sim$ 400 cm$^{-3}$, which possibly explains the asymmetric apparent shape.

\subsection{\textit{Fermi}-LAT Spectral Energy Distribution}
For the analysis of the source spectrum, we use the LAT data in the energy range between 1\,GeV and 1\,TeV. The photon events are divided in 18 logarithmically distributed energy bins. 
The diffuse background, as well all the sources within 2$^{\circ}$ of the center position of the PWN, are also fitted during the spectral analysis.
The spectral energy distribution (SED) is fitted with a LogParabola model as well as with a Broken Power Law (Broken PL) model in order to better estimate the break energy. The function used for the Broken PL law is

\begin{equation}
\label{broken_pl}
\frac{dN}{dE} = N_{0} \times 
\begin{cases} 
	(\frac{E}{E_{b}})^{- \Gamma_{1}} \; \mathrm{if} \: E<E_{b};
	\\
	(\frac{E}{E_{b}})^{- \Gamma_{2}} \; \mathrm{otherwise} \:.
\end{cases}
\end{equation}

In order to measure the spectra of the PWN, we perform the spectral analysis twice: 
after a preliminary optimization of the parameters of the sources included in the model, the spectral analysis is performed for the first time in the whole energy range 1\,GeV-1\,TeV. 
We perform the spectral analysis a second time in the various energy ranges using the resulting templates for the PWN morphology obtained from the localization and extension analysis in each energy bin (see Section \ref{ext_2DGaussian}).

The fit results for the LogParabola and Broken PL are reported in the ``\textit{Fermi}'' column of Table \ref{table_sed}.
The ``\textit{Fermi} + H.E.S.S.'' column of the Table \ref{table_sed} contains the resulting spectral information for the PWN HESS J1825-137 from a joint fit to the recent H.E.S.S. data \cite{2019A&A...621A.116H} and energy flux derived in this work with \textit{Fermi}-LAT.

\begin{table*}[h]
\small
\centering
\begin{tabular}{ccc|ccc}
LogParabola & & & Broken PL & & \\
\hline
Parameter &  \textit{Fermi} &  \textit{Fermi} $+$ H.E.S.S. & Parameter &  \textit{Fermi} &  \textit{Fermi} $+$ H.E.S.S. \\
\hline
\hline
$\alpha$ &  2.05 $\pm$ 0.18 & 2.13 $\pm$ 0.03 & $\Gamma_{1}$ & 1.84 $\pm$ 0.04 & 1.88  $\pm$ 0.03\\
$\beta$ &  0.040  $\pm$ 0.013 & 0.061 $\pm$ 0.002 & $\Gamma_{2}$ &  2.08 $\pm$ 0.06 & 2.51  $\pm$ 0.01\\
$E_{0}$ (GeV) &  79 $\pm$ 18 & 78 $\pm$ 18 & $E_{b}$ (GeV) & 26 $\pm$ 1 & 139 $\pm$ 17 \\
$N_{0}$  & 6.41 $\pm$ 0.73 & 5.99 $\pm$ 0.21 & $N_{0}$  & 6.79 $\pm$  0.41 & 7.55 $\pm$  0.46\\
$\chi^{2}$/ndf & 9/14 & 29/28 & $\chi^{2}$/ndf & 10/14 & 81/28\\
\hline
\end{tabular}
\caption{\small \label{table_sed}
Fit parameters for the SED of HESS J1825-137 with a LogParabola function and with a Broken PL function. The scale of the parameter normalization $N_{0}$ is ($10^{-11}$ erg cm$^{-2}$ s$^{-1}$). The ``\textit{Fermi}'' and ``\textit{Fermi} $+$ H.E.S.S.'' columns contain the fit results obtained with only the \textit{Fermi}-LAT flux and with both \textit{Fermi}-LAT and H.E.S.S. fluxes respectively.}
\end{table*}

\begin{figure}[h]
\centering
\includegraphics[scale=0.69]{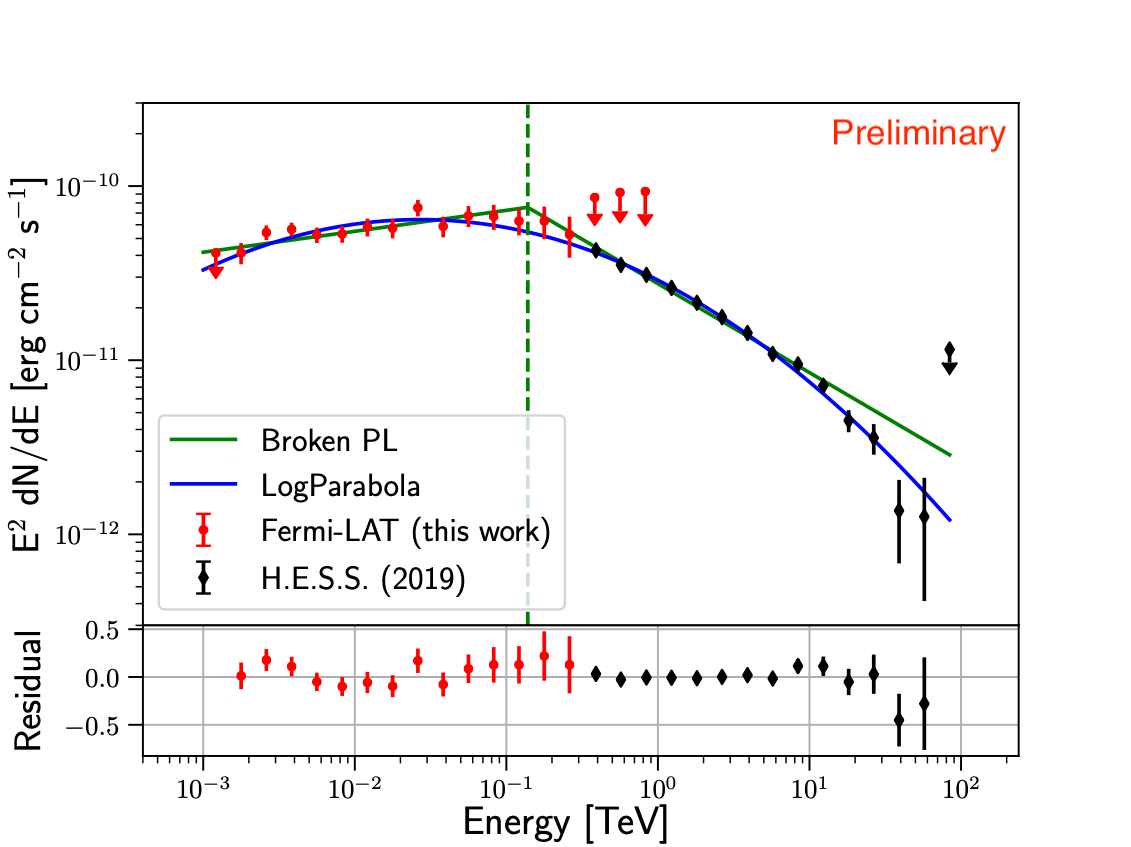}
\caption{\small \label{spectra_hess_j1825}
Combined spectra of the PWN HESS J1825-137 with the spectral measurements obtained in this work (red points) using 10 years of \textit{Fermi}-LAT data from 1\,GeV to 1\,TeV and the H.E.S.S. results for the 100\,GeV -- 90\,TeV band (black points). The combined SED has been fitted with both a LogParabola (blue line) and a Broken PL (green line). The vertical line corresponds to the break energy $E_{b}$ of the Broken PL. The bottom panel shows the normalized residual between data and the LogParabola. 
}
\end{figure}

Figure \ref{spectra_hess_j1825} shows the combined SED with the results obtained in this work, using 10 years of \textit{Fermi}-LAT
for the 1\,GeV -- 1\,TeV energy band, and the H.E.S.S. results for the 100\,GeV -- 90\,TeV energy range \cite{2019A&A...621A.116H}.

\subsection{Energy dependent analysis of HESS\,J1825-137}
\label{ext_2DGaussian}
For the analysis of the energy-dependent extension the spectra of HESS\,J1825-137 and of the nearby sources, have been fixed to the Table \ref{table_sed} values. The normalization values of the isotropic and diffuse components are instead left as free parameters of the fit. 

Before performing the extension analysis, the localization is again optimized inside each energy bin. The localization and extension analysis are performed in this part separately in each energy bin. We use 2 bins per decade in the 1--100\,GeV band and a single bin in the 100\,GeV\,--\,1\,TeV energy band. A variation of the PWN position between the different bins is observed. Moving from the lowest energy bin, 1--3\,GeV, to the highest energy one, 100\,GeV -- 1\,TeV, the fitted 2DGaussian centroid of the PWN moves toward the position of the pulsar.
The extension analysis is performed by fitting a 2D-Gaussian template in each energy bin.

\begin{figure*}[h]
\begin{minipage}{.5\textwidth}
\includegraphics[scale=0.55]{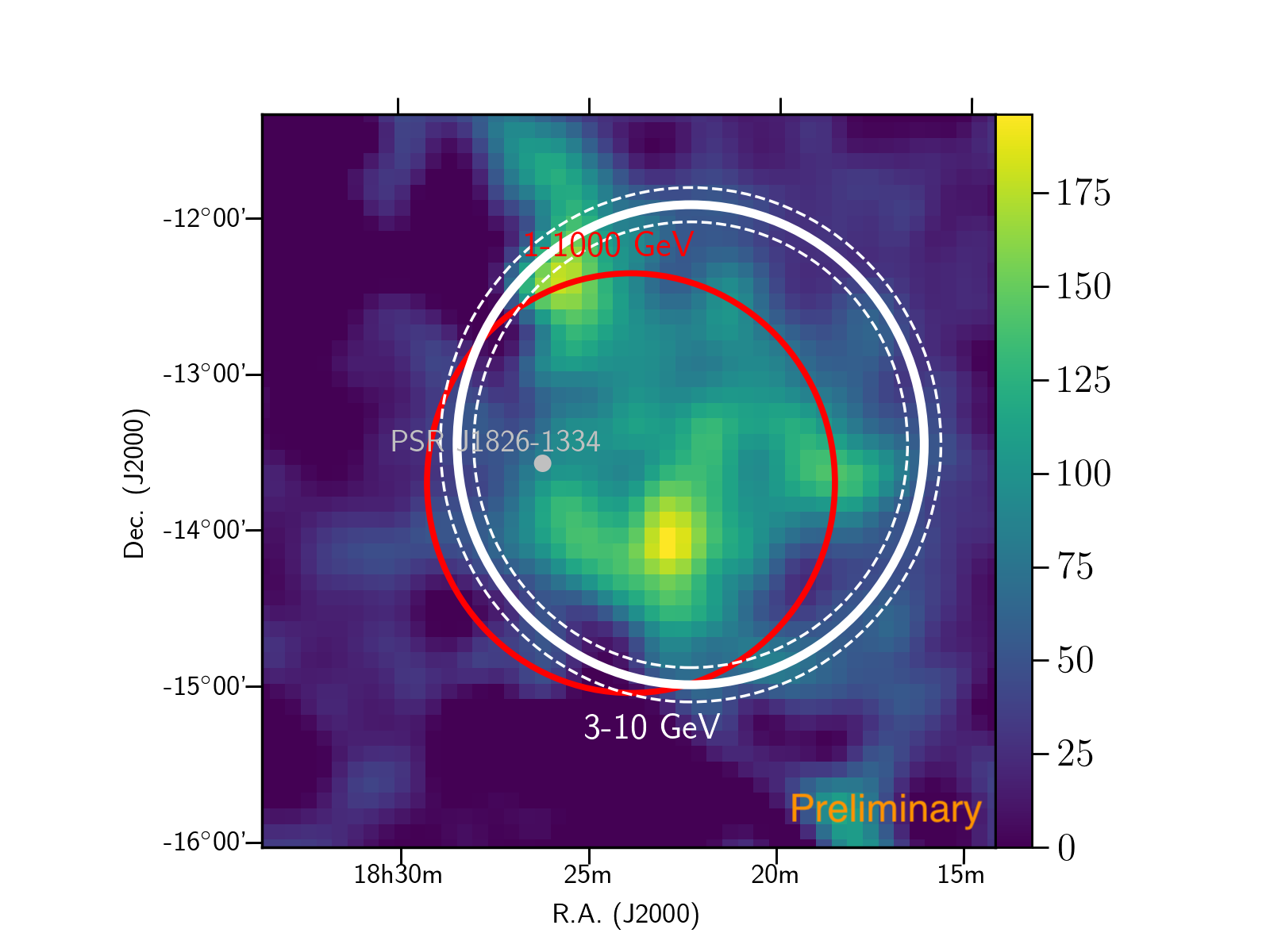}
\end{minipage}%
\begin{minipage}{.5\textwidth}
\includegraphics[scale=0.55]{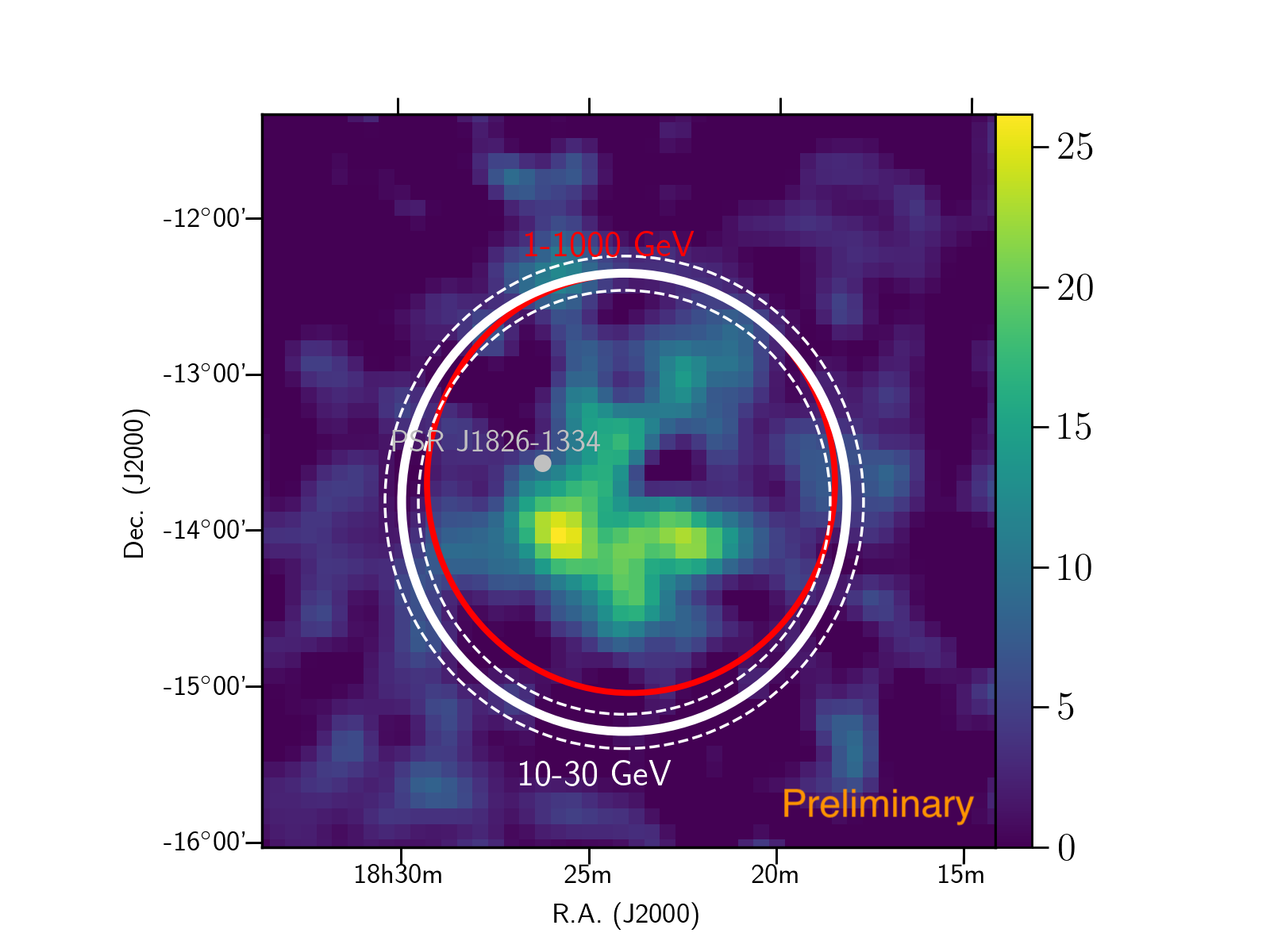}
\end{minipage}%
\\
\begin{minipage}{.5\textwidth}
\includegraphics[scale=0.55]{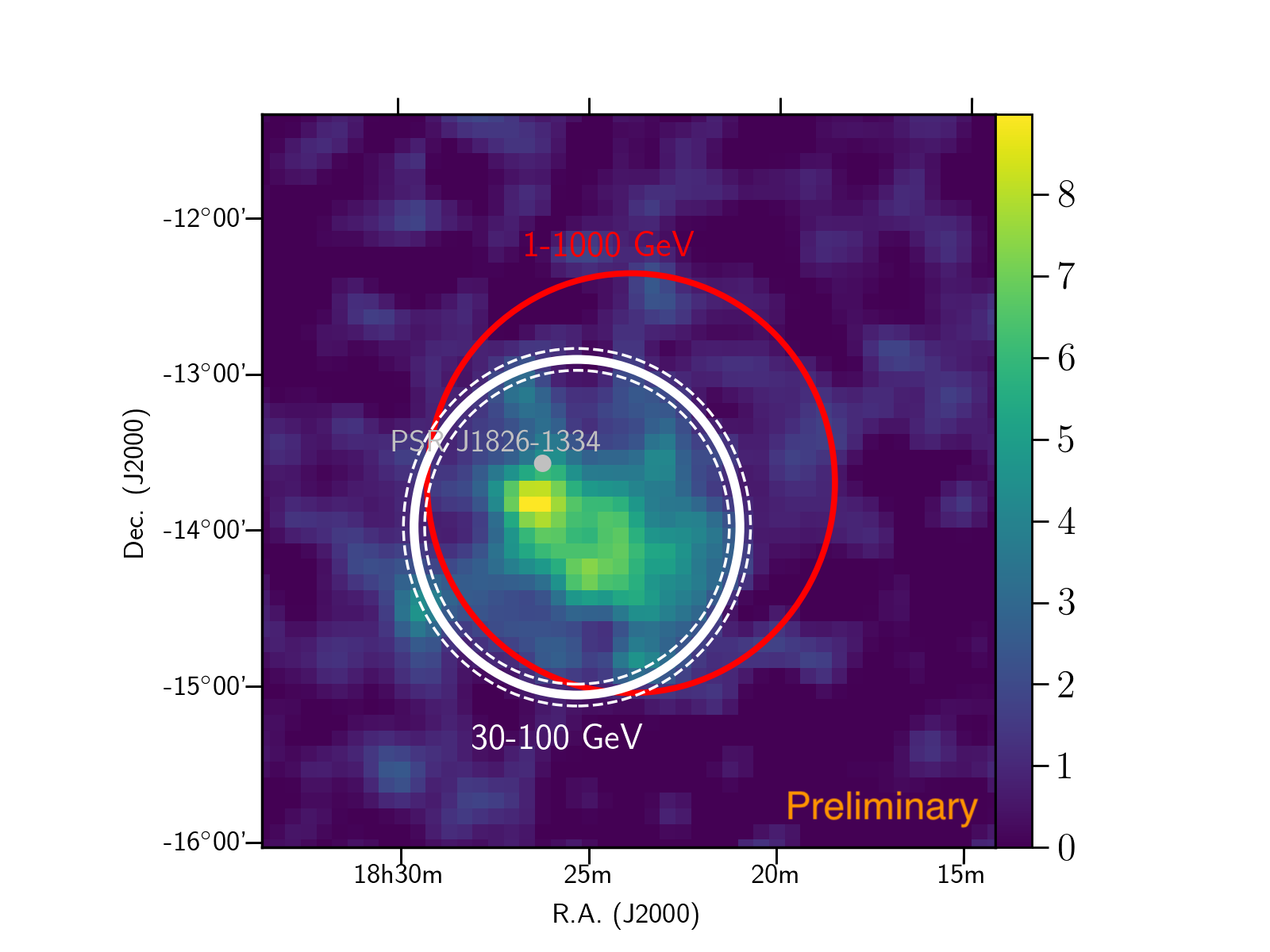}
\end{minipage}%
\begin{minipage}{.5\textwidth}
\includegraphics[scale=0.55]{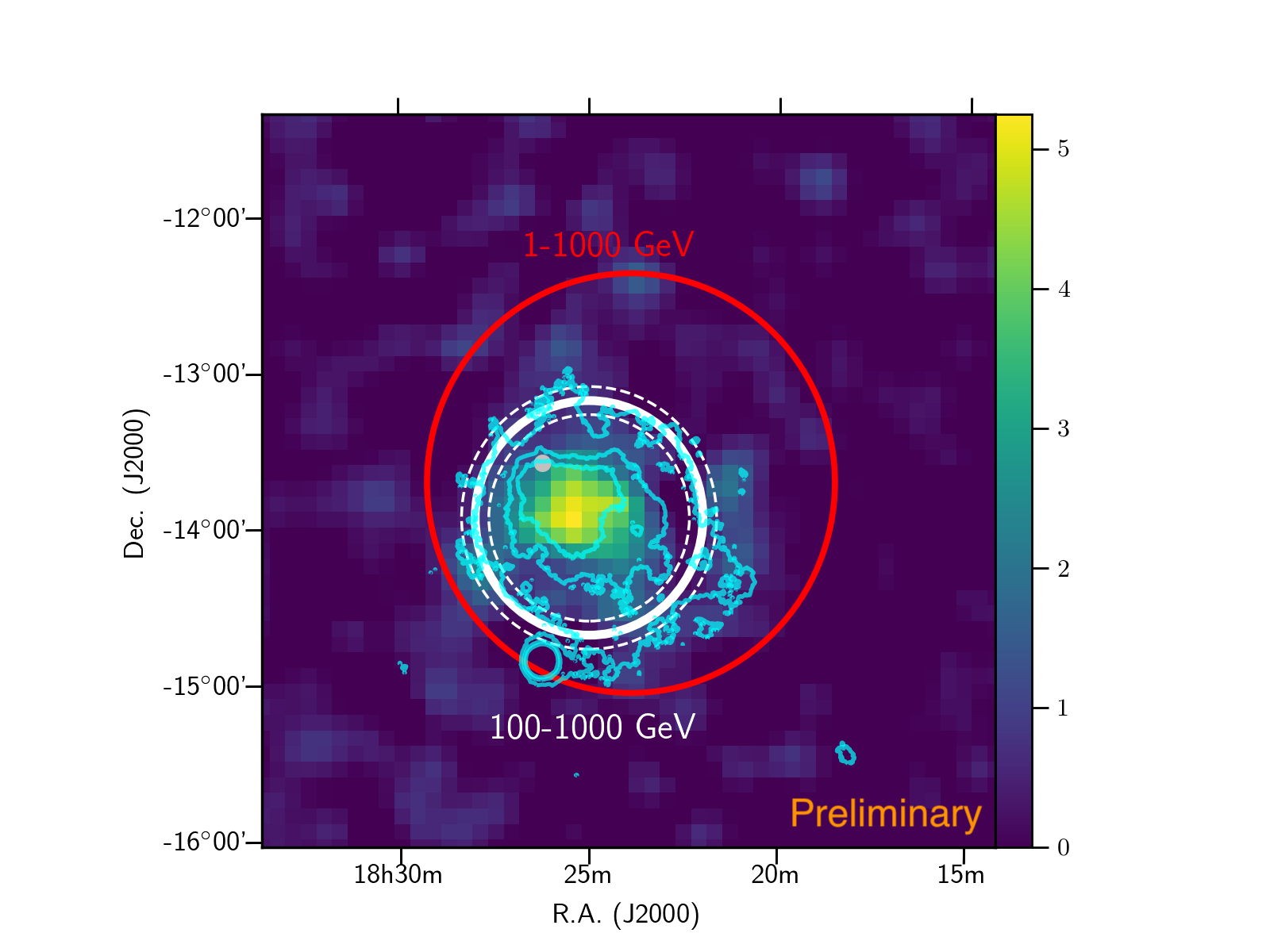}
\end{minipage}%

\caption{\small \label{significance_maps}  Excess maps, in celestial coordinates, of the region around HESS\,J1825-137 for the energy ranges: 3--10\,GeV (top left), 10--32\,GeV (top right), 32--100\,GeV (bottom left) and 100\,GeV -- 1\,TeV (bottom right). In the energy bin between 1 and 3\,GeV the source is not completely resolved from diffuse emission (TS$<$25).
The excess maps are smoothed with a Gaussian of radius 0.$^{\circ}$1. The white circles represent the extension (solid line) and its uncertainty (dashed lines) determined in the single energy bin.
For a comparison the resulting extension obtained for the entire energy range (1\,GeV -- 1\,TeV) is plotted with a red line. All the extensions shown correspond to the 68\% containment radius. In the 100\,GeV -- 1\,TeV (bottom right) plot, H.E.S.S. significance contours at 5, 10, and 15 $\sigma$, for energies below 1 TeV \cite{2019A&A...621A.116H}, are shown with light-blue lines for a comparison. 
}
\end{figure*}

In the energy band between 1 and 3\,GeV the source is not resolved,
therefore no significant result is reported for the extension below 3 GeV. 
This is connected to the reduced source flux at low energy, as well as the large influence of the diffuse emission in this energy range which does not allow the source to be easily distinguished from the background.
Figure \ref{significance_maps} presents the source excess maps for each energy bin used in the energy-extent analysis.

Table \ref{table_extent} contains the results of the extent measurements performed with the 2D-Gaussian template in 2 bins per decade in the 3--100\,GeV band and in a single bin for the 100\,GeV\,--\,1\,TeV band. With the extension is reported also the total error. For the estimation of systematic error $\sigma_{syst}$, the analysis has been repeated using two different diffuse models: $\sigma_{syst} = \sqrt{(\mathrm{Ext}_{D1}-\mathrm{Ext}_{D2})^2 + (\sigma_{Ext_{D1}} - \sigma_{Ext_{D2}})^2}$. 

\begin{table*}[h]
\small
\centering
\begin{tabular}{ccc}
Energy range (GeV) &  Extension ($^{\circ}$ ) & TS$_{ext}$ \\
\hline
\hline
3 -- 10 & $1.54 \pm 0.11$ & 219 \\ 
10 -- 32 & $1.47 \pm 0.11$ & 234 \\
32 -- 100 & $1.07 \pm 0.07$ & 207 \\
100 -- 1000 & $0.75 \pm 0.09$ & 129 \\
\hline
\end{tabular}
\caption{\small \label{table_extent}
Extension measurements as a function of the energy with statistical and systematic errors. The extension is characterized by the 68\% containment radius obtained from a 2D-Gaussian template fit. The systematic error $\sigma_{syst}$, due to the diffuse model, has been calculated as $\sigma_{syst} = \sqrt{(\mathrm{Ext}_{D1}-\mathrm{Ext}_{D2})^2 + (\sigma_{Ext_{D1}} - \sigma_{Ext_{D2}})^2}$.}
\end{table*}

\section{Conclusions}
We have analyzed 10 years of \textit{Fermi}-LAT data in the energy range between 1 GeV and 1 TeV and we have performed, for the first time with \textit{Fermi}-LAT data, an energy dependent analysis of the extension of the PWN HESS\,J1825-137.
The nebula HESS J1825-137 presents a strong energy dependent morphology. The spectral index was seen by \cite{2006A&A...460..365A} to soften with increasing distance from the PSR, this implies that the population of electrons in the nebula had travelled and cooled out to large distances. 
The fact that the low-energy electrons at large distances from the pulsar produce the softest spectrum was interpreted to mean that these are the oldest electrons in the system. In this work, the observation was extended low to few GeV confirming the emission at large distances by the lowest energetic particles.

Although a variation of the nebula extension is expected at TeV energy, a similar change is not expected at \textit{Fermi}-LAT energies because the age of the PWN  is smaller than the typical time of energy losses. 
We found continued morphological changes and increasing size of this PWN towards lower energies, which would be difficult to explain with a simple diffusion-energy loss process.

\subsection*{ACKNOWLEDGMENTS}
The \textit{Fermi}-LAT Collaboration acknowledges support from NASA and DOE (United States), CEA/Irfu, IN2P3/CNRS, and CNES (France), ASI, INFN, and INAF (Italy), MEXT, KEK, and JAXA (Japan), and the K.A. Wallenberg Foundation, the Swedish Research Council, and the National Space Board (Sweden).

\end{document}